# Title

A phase transition in monetary function explains expansion without inflation

# Author

Ran Huang[1]
[1]Academy for Small Commodity Economy, Yiwu Research Institute of Fudan University, Yiwu 32200, China

huangran@fudan.edu.cn

# Abstract

Large monetary expansions do not necessarily generate consumer-price inflation, challenging scalar views of "money supply." Here we propose that monetary function is phase-dependent: newly issued base money can occupy distinct functional compartments with different coupling to prices. Starting from an accounting framework that separates reproduction, consumption, and reservation, we operationalize a measurable order parameter, $\phi = RB/MB$, the reserve-share fraction of the monetary base. Using Japan's monthly record (1971–2026), we identify a compositional phase transition after 2013 from a cash-dominated to a reserve-dominated regime, quantitatively captured by a Landau-type order-parameter transition. Phase-conditional local projections using unexpected (residual) base-growth shocks show that, in Japan, unexpected base expansions are absorbed primarily as reserve balances—$\phi$ rises significantly—rather than entering the consumption-goods transaction sector; consequently, the core CPI inflation response is strongly attenuated and can even reverse sign. This demonstrates that increases in monetary supply do not necessarily cause inflation: the key is the "phase" in which incremental money accumulates (reservoir versus circulation). We further define function-specific efficiencies for reservation absorption and CPI transmission and provide an operational distinction between circulation-driven and reservation-dominant inflation regimes.

# Monetary expansion without inflation: a missing state variable

Monetary expansions are often assumed to scale into higher prices. Yet modern episodes repeatedly violate this intuition: central bank balance sheets can grow dramatically while consumer-price inflation remains subdued for years.

Conventional explanations invoke changes in velocity, expectations, globalization, or sectoral slack, but they typically treat "money" as a scalar quantity whose effect is modulated by auxiliary factors. This framing obscures a more basic possibility: the effect of monetary issuance may depend on where newly created nominal claims reside within the monetary system.

We propose a simple organizing principle: the effect of monetary issuance depends on the phase of money—that is, on the allocation of base money across functional compartments. In this view, monetary injections can populate a transactional, cash-like compartment or a reservation reservoir (settlement liquidity held as reserves), and these compartments have different effective coupling to consumer prices. The central empirical question therefore becomes not only "how much money is issued," but also "in which phase does it accumulate, and how does the transmission kernel change across phases?"

Japan provides a uniquely long and well-measured record for testing this idea.[1,2] From the early 1970s to the mid-2020s, Japan spans high-inflation decades, the post-1990 low-inflation era, and the post-2013 balance-sheet expansion period, while the monetary base series explicitly decomposes into cash and reserve balances (Bank of Japan statistics). [3,4] This allows us to define an order parameter for monetary phase and to test whether inflation transmission changes across phases.

## A minimal accounting theory: SCR, SRF priority, and two inflation regimes

We start from a minimal accounting model of real economic output $G_t$ allocated into three functional uses:

$$G_t = S_t + C_t + R_t \quad (1)$$

where $S_t$ supports reproduction (seed investment), $C_t$ is contemporaneous consumption, and $R_t$ is reservation—resources held as buffers against uncertainty and for intertemporal transfer. The key point is functional rather than commodity-specific: societies maintain a reservation stock because uncertainty and timing mismatch make purely "just-in-time" consumption infeasible.

In the original formulation, reservation assets are subject to effective loss—conceptually described as deterioration or "rotting." A symbolic instrument $E_t$ (money) becomes valuable because it can replace part of the reservation stock with a more durable and transferable nominal form—termed the Saving-Replacement Function (SRF). This perspective implies a priority ordering: when new nominal claims are issued, they may be absorbed first as reservation buffers rather than immediately amplifying consumption demand. The SCR/SRF framework and its inflation typology are developed in full in the original preprint. [5]

In modern economies, reservation assets need not literally decay; instead, they face effective carrying costs and frictions—opportunity costs, regulatory liquidity demand, intermediation constraints, and institutional incentives—that govern how liquidity is stored and mobilized. The fundamental mechanism remains the same: reservation is a distinct compartment with its own dynamics and constraints. This motivates two operational inflation regimes consistent with the original distinction:

- **Circulation-driven inflation (Type I):** nominal claims enter the transactional or consumption compartment, raising consumer prices when supply-side adjustment is limited.
- **Reservation-dominant regime (Type II):** nominal claims accumulate as reservation buffers, changing balance-sheet composition and weakening CPI transmission unless reservoir liquidity is released into circulation; price effects may also appear outside CPI, for example in asset prices, yields, or risk premia.

To test this framework in modern data, we need a measurable state variable capturing the phase allocation of base money.

## Order parameter for the phase of money

In contemporary monetary systems, the monetary base is the central bank's highest-grade nominal liability. Crucially, it is partitioned into cash in circulation and reserve balances. We therefore define an order parameter:

$$\phi_t \equiv \frac{RB_t}{MB_t} \quad (2)$$

where $MB_t$ is the monetary base (average outstanding) and $RB_t$ is reserve balances. $\phi_t$ measures the fraction of base money residing in the reservation reservoir. Low $\phi$ corresponds to a cash-dominated regime ("cash-phase"), while high $\phi$ corresponds to a reserve-dominated regime ("reserve-phase").

This mapping is conservative: it does not require observing $S_t$, $C_t$, and $R_t$ directly. It only requires that reserve balances behave as a reservation stock in the modern balance-sheet sense, and that consumer-price inflation reflects the state of the consumption sector. Under SRF priority, a core prediction is **reservoir trapping**: unexpected base expansions should raise $\phi$, indicating that issuance is absorbed into reservation rather than into transactions.

# A minimal physical model of phase-dependent monetary transmission

To show that the physical framing is not merely descriptive, we introduce a minimal calculable model for phase-dependent monetary transmission. The purpose of this section is not to reproduce the full technical development, which is presented in Supplementary Note 2, but to state the smallest dynamical structure needed for the empirical analysis below.

We represent the system by two effective compartments: a reservation reservoir $R$ and a circulation-coupled liquidity pool $X$. The first absorbs balance-sheet buffering and liquidity storage; the second remains coupled to consumer-price formation. The physical premise is that reservation is a dissipative channel, so incremental symbolic capacity need not be transmitted directly into the consumption sector.

The starting point is the allocation identity $G_t = S_t + C_t + R_t$, together with the minimal post-shock dynamics

$$\dot{R} = -\delta_R R + \mu X, \dot{X} = -\gamma X \quad (3)$$

where $\delta_R$ is the effective relaxation rate of the reservoir, $\gamma$ is the decay rate of circulation-coupled liquidity, and $\eta$ is the reabsorption flow from $X$ back into $R$.

In this way we have the order parameter

$$\phi = \frac{R}{R+X} = \frac{RB}{MB} \quad (4)$$

which measures the share of nominal capacity residing in the reservation reservoir.

To connect the model to prices, we introduce a phase-dependent CPI coupling law,

$$\Delta\pi = \chi(\phi)X \quad (5)$$

with a critical threshold $\phi_c$ such that CPI transmission is positive below the threshold and attenuated or reversed above it. The detailed constitutive form and its calibration are given in Supplementary Note 2; the essential implication for the main text is that the cash phase and reserve phase are interpreted as two sides of a single phase-conditioned transmission mechanism rather than as two unrelated regimes.

Under this formulation, the empirical analysis below has two direct tasks: first, to identify whether Japan exhibits an observable transition in the order parameter; and second, to test whether shock absorption and CPI transmission behave in the direction predicted by the model. The role of the main text is therefore to validate the phase structure empirically, while the full derivation, critical-point construction, Landau coarse-graining, and supplementary simulations are deferred to the Supplementary Materials.

## Japan: a compositional phase transition after 2013

We use monthly Japanese data (1971–2026) comprising BOJ monetary base components and CPI (headline and core; all items less fresh food and energy). [1,2,6,7] Figure 1 shows two facts side-by-side. First, the monetary base increases by orders of magnitude after 2013, whereas CPI evolves much more smoothly (Fig. 1A). Second, the order parameter $\phi$ undergoes a pronounced upward shift around 2013, indicating a transition from cash-dominance to reserve-dominance (Fig. 1B).

We quantify this compositional transition using an order-parameter transition form familiar in statistical physics:

$$\phi(t) = \phi_0 + A \tanh\left(\frac{t-t_0}{w}\right) \quad (6)$$

The fit over 2010–2018 yields a well-defined critical month $t_0$ near 2013 and a finite transition width $w$ (Fig. 2), consistent with a genuine compositional phase change rather than a gradual drift. Breakpoint detection across alternative windows is robust and places the core compositional change in the same period (Fig. S1).

This establishes a first conclusion: Japan's post-2013 expansion is not merely "more money," but money in a different phase.

## Phase diagram: reserve dominance constrains transmission rather than uniquely determining inflation

Figure 3 plots the monthly observations in the $(\phi, \pi^{core})$ plane, where $\phi$ is the reserve-share order parameter and $\pi^{core}$ is core CPI inflation. The figure serves a specific interpretive purpose: it shows that reserve dominance should not be read as a one-to-one predictor of inflation, but rather as a marker of the **transmission regime** in which monetary shocks are processed.

The key empirical fact is that high-$\phi$ states are compatible with more than one inflation outcome. In Japan, the reserve-dominated phase includes both the near-zero inflation period of 2013–2021 and the subsequent inflation recovery after 2022. This means that the order parameter does not mechanically determine the level of CPI inflation. Instead, it constrains how strongly monetary injections couple into the price-forming sector.

This is precisely the role assigned to $\phi$ in the physical model. The model does not imply that reserve dominance eliminates inflation under all conditions. Rather, it implies that the effective CPI transmission kernel becomes phase-dependent, so that once the system moves into a high-$\phi$ regime, the relationship between monetary expansion and consumer prices is no longer governed by the same coupling structure that prevailed in the cash-dominated phase.

Figure 3 therefore occupies an important middle position in the logic of the paper. Figures 1 and 2 establish that Japan experienced a compositional phase transition; Figure 3 shows that this transition relocates the system into a different region of state space, one in which inflation outcomes remain possible but are governed by a different transmission structure. This prepares the ground for the dynamic tests that follow.

In this sense, the phase diagram should be interpreted not as a static claim that "high $\phi$ means low inflation," but as evidence that reserve dominance changes the geometry of monetary transmission. The relevant question is not whether inflation can occur in the reserve phase, but whether the system remains on the same response surface once the order parameter has crossed into a different regime. Figure 3 shows that it does not.

# Phase-dependent inflation kernel and reservoir trapping

A phase transition in composition would matter only if it changes dynamics. We therefore estimate impulse responses using phase-conditional local projections, separating the sample into cash-phase ($\phi < 0.30$) and reserve-phase ($\phi > 0.60$), while excluding the intermediate critical region. Local projections follow Jordà. [8] To mitigate endogeneity from policy reactions, we construct unexpected base-growth shocks as within-phase AR(12) residuals of the year-on-year growth of seasonally adjusted base money, standardized within each phase (full specifications and robustness checks are provided in the Supplementary Materials).

Figure 4 reports the phase-conditional impulse responses of core CPI inflation to an unexpected base-growth shock. The response is weakly positive in the cash-phase, consistent with Type I circulation dynamics. Strikingly, the reserve-phase exhibits a significantly negative medium-horizon response. The sign and shape of the inflation kernel therefore depend on phase: the reserve-dominant system does not behave as a scaled version of the cash-dominant system.

To close the mechanism loop, we test whether the same shocks preferentially populate the reservation reservoir. Figure 5 reports the impulse response of $\phi$ to the unexpected base-growth shock. In both phases, $\phi$ rises significantly at medium horizons, with smaller amplitude in the reserve-phase, consistent with saturation near a high-$\phi$ plateau. This is direct evidence of reservoir trapping: unexpected expansions of base money increase the reserve-share fraction, indicating that issuance is absorbed into the reservation compartment.

Together, Figs. 4–5 support a coherent mechanism consistent with the SCR/SRF bridge: base money expansions populate reservation first ($\phi \uparrow$), and CPI inflation transmission is consequently muted or altered in the reserve-dominated phase. This operationalizes the SRF priority emphasized in the original SCR framework.

## Functional efficiencies of monetary issuance

The original efficiency intuition for currency can be modernized from intrinsic production cost to functional performance. Using impulse responses, we define:

$$Eff^{(R)} \equiv \max_{h \leq H} \left|\frac{\partial \phi_{t+h}}{\partial u_t^{MB}}\right|, \qquad Eff^{(C)} \equiv \max_{h \leq H} \left|\frac{\partial \pi_{t+h}^{core}}{\partial u_t^{MB}}\right| \quad (7)$$

where $u_t^{MB}$ is the unexpected base-growth shock. $Eff^{(R)}$ captures efficiency at filling reservation buffers (SRF absorption), while $Eff^{(C)}$ captures efficiency at generating CPI pressure. Japan demonstrates that an economy can be efficient in SRF absorption, as shown by strong reservoir trapping, while being inefficient—or qualitatively altered—in CPI transmission in the reserve-phase.

## Implications, scope, and limitations

Our results provide a testable distinction between two inflation regimes: a circulation-driven regime in which monetary shocks couple positively into consumer prices, and a reservation-dominant regime in which issuance is trapped in a reserve reservoir and CPI coupling is weakened or altered. The Japan record shows that large monetary expansions can coincide with low CPI inflation not because money is generically "ineffective," but because it occupies a phase whose primary function is reservation.

The framework is intentionally minimal and empirically testable. In the present paper, we establish three core claims. First, $\phi$ behaves as an order parameter for a compositional transition in the monetary base. Second, unexpected base expansions increase $\phi$, consistent with reservoir trapping. Third, the CPI response kernel is phase-dependent, indicating that the reserve-dominant system does not behave as a scaled version of the cash-dominated system.

These claims are deliberately narrow. We do not argue that $\phi$ alone determines inflation, nor that the reserve phase must always imply negative CPI responses. Rather, $\phi$ identifies a transmission regime in which CPI coupling can be weak, altered, or condition-dependent unless additional factors enable reservoir-to-circulation release. Nor do we require that all price effects appear in CPI: reservation-dominant monetary expansions may also operate through other channels, including asset prices, yields, or risk premia. These extensions are natural next steps, but they are not required for the validity of the mechanism established here.

## Discussion: first-principles interpretation of phase-dependent monetary transmission

Our results reorganize the "expansion without inflation" puzzle around a missing state variable: the phase allocation of base money. Once the monetary base is treated as a partitioned stock whose components have distinct functions and distinct couplings to prices, the Japan case becomes structurally intelligible rather than anomalous.

The first-principles bridge to the SCR/SRF framework begins from the conservation identity in Eq. (1), in which output is allocated across reproduction, contemporaneous consumption, and reservation. In the original formulation, the key physical ingredient is that reservation is a dissipative compartment. Whether interpreted as literal deterioration or, in a modern balance-sheet setting, as effective carrying costs, liquidity regulation, precautionary absorption, and intermediation friction, reservation is the compartment in which losses, constraints, and buffer requirements accumulate. In such a system, an incremental symbolic input is not generically converted into "work" in the consumption channel. It is first drawn into the compartment that compensates losses and relaxes constraints. This is the physical meaning of SRF priority.

Japan provides a modern balance-sheet instantiation of this logic. The order parameter $\phi$ defined in Eq. (2) operationalizes the reservation share of the monetary base: reserve balances represent the highest-grade settlement buffer, while CPI inflation proxies the price state of the consumption compartment. The empirical mechanism loop closes at precisely the point where SCR/SRF predicts it should. Unexpected base expansions significantly increase $\phi$, indicating that incremental issuance is first absorbed into reservation rather than immediately coupling into consumer prices.

The Japan evidence validates three structural components of the SCR/SRF view. First, SRF priority becomes empirically observable: the first destination of new symbolic capacity is reservation rather than immediate consumer-price transmission. Second, phase transition appears as a change in the dominant compartment. The sharp post-2013 shift in $\phi$, quantified by the Landau-type transition in Eq. (5), indicates that the monetary base moves into a reserve-dominated configuration. Third, the two inflation regimes become operationally identifiable through response kernels. Type I corresponds to a regime in which monetary shocks couple positively into consumer prices, whereas Type II corresponds to a reservation-dominant regime in which CPI coupling is weakened or altered because shocks are trapped in reservation. Japan exhibits precisely this phase dependence: the inflation response kernel differs markedly between the cash phase and the reserve phase, and this difference remains after residualizing the monetary shock.

This interpretation also clarifies what the Japan case does and does not show. "Money supply does not necessarily cause CPI inflation" is not a paradox once issuance is understood as symbolic capacity that can be allocated across compartments. CPI inflation is a property of the consumption channel, whereas monetary issuance adds nominal capacity that may first be absorbed by reservation. When the reservation share rises, marginal issuance can be used primarily to fill or stabilize buffer stocks rather than to intensify transactions in the price-forming sector. Weak CPI transmission in this setting is therefore not evidence of generic monetary ineffectiveness, but of functional reallocation toward reservation.

This point is further clarified by the functional efficiencies defined in Eq. (7). Japan illustrates a regime in which reservation absorption remains robust, because unexpected base expansions are efficiently absorbed into the reservation reservoir, whereas CPI transmission is muted or qualitatively altered in the reserve phase. A monetary action may therefore be ineffective at generating CPI inflation while still being highly effective at performing reservation replacement and balance-sheet stabilization.

In this sense, the Japan case does not merely offer another example of quantitative easing with unexpectedly low inflation. It provides a compact empirical validation of a phase-based monetary mechanism. Once the relevant state variable is identified, expansion without inflation ceases to be an anomaly and becomes the expected outcome of a reserve-dominated system in which newly created nominal capacity is first absorbed by a dissipative reservation reservoir rather than by the circulation-coupled price-forming sector.

# Conclusion

We propose that monetary issuance is phase-dependent: base money can occupy distinct functional compartments with different coupling to consumer prices. Operationalizing this idea with an order parameter $\phi = RB/MB$. Japan's monthly record (1971–2026) reveals a sharp post-2013 compositional phase transition from a cash-dominated to a reserve-dominated regime, well quantified by a Landau-type order-parameter transition.

Phase-conditional local projections using unexpected (residual) base-growth shocks show two linked facts: shocks raise $\phi$ (reservoir trapping), and the inflation response kernel differs across phases, with weak positive coupling in the cash-phase but significantly altered—and here negative at medium horizons—coupling in the reserve-phase. This closes a compact empirical mechanism loop explaining

"expansion without inflation": incremental issuance is first absorbed by reservation rather than circulation.

The findings connect directly to first-principles elements of the SCR/SRF framework—conservation of allocation with a dissipative reservation channel—in which incremental symbolic capacity is expected to compensate losses and relax constraints before generating consumption pressure. More broadly, the results suggest that the key question is not only how much money is issued, but in which phase it accumulates, providing a testable basis for cross-country evaluation of phase-conditioned monetary transmission.

# Acknowledgment

This work is financially supported by the YRI-FD Industrial Project (YRI-IP-25-01).

# Figures

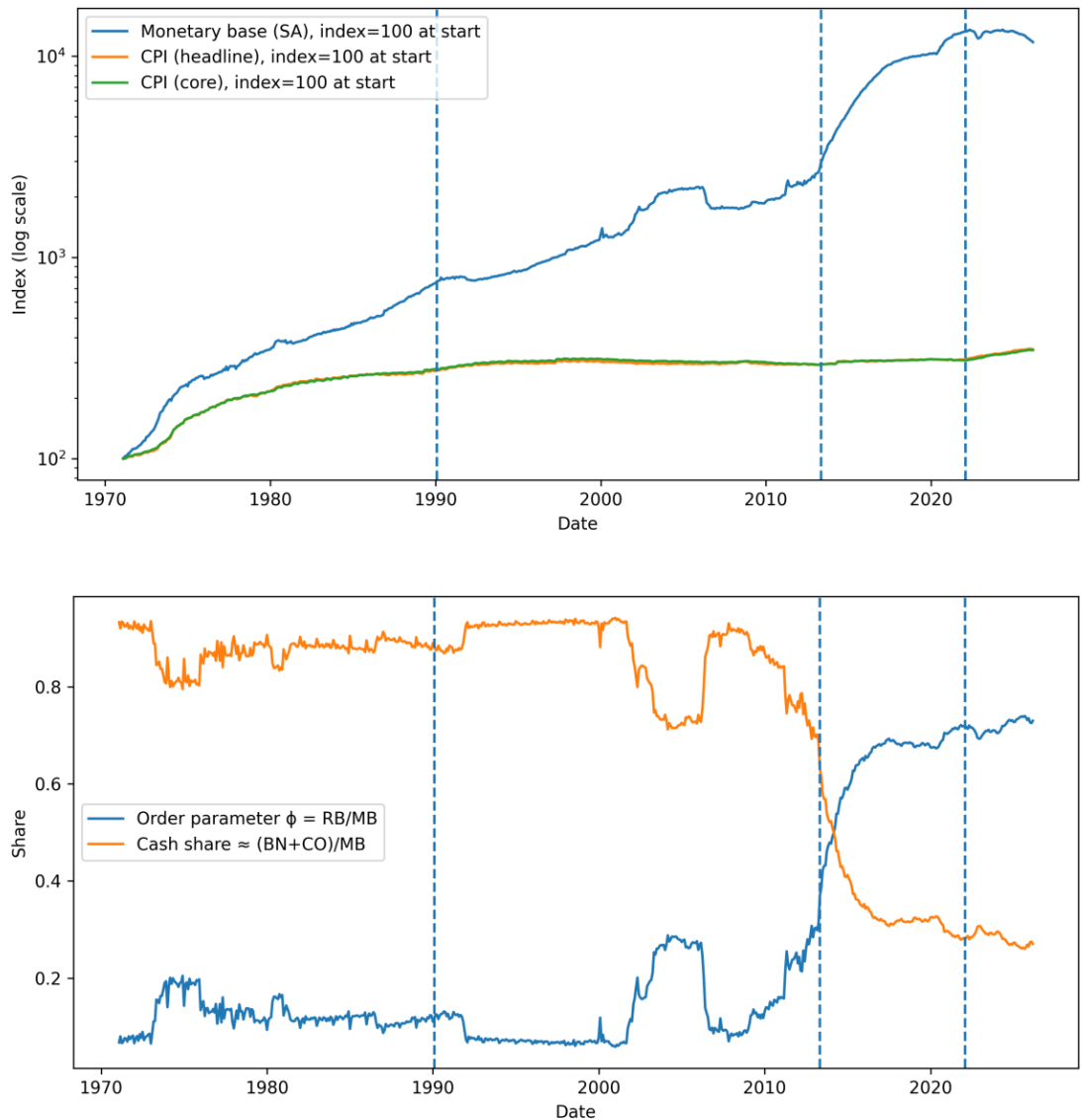

**Fig. 1. Monetary base expansion, CPI, and the compositional shift in the phase of money in Japan (1971–2026).**
(A) Indexed trajectories (start = 100; log scale) of the seasonally adjusted monetary base ($MB^{SA}$), headline CPI, and core CPI (all items less fresh food and energy). (B) Order parameter $\phi = RB/MB$ (the reserve-share fraction of the monetary base) and the corresponding cash share, $1 - \phi$. Vertical dashed lines mark 1990-01, 2013-04, and 2022-01. Monetary base data are from the BOJ long-term time series; CPI data are from Japan official statistics (2020 base).

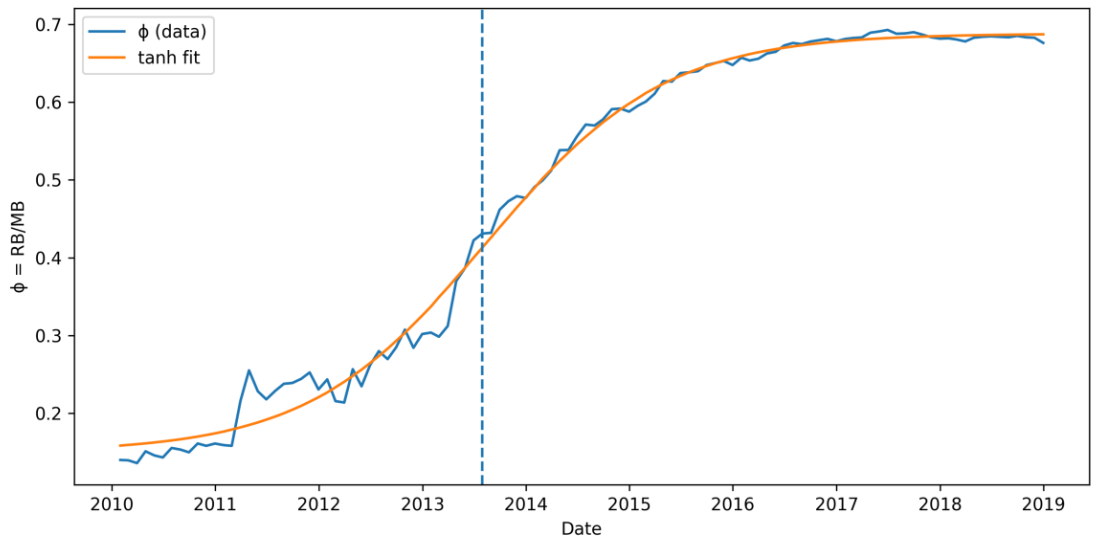

**Fig. 2. Landau-type order-parameter transition after 2013.**
Order parameter $\phi$ over 2010–2018 together with the nonlinear fit $\phi(t) = \phi_0 + A\tanh((t-t_0)/w)$. The dashed line indicates the fitted critical month $t_0$, and $w$ denotes the transition width.

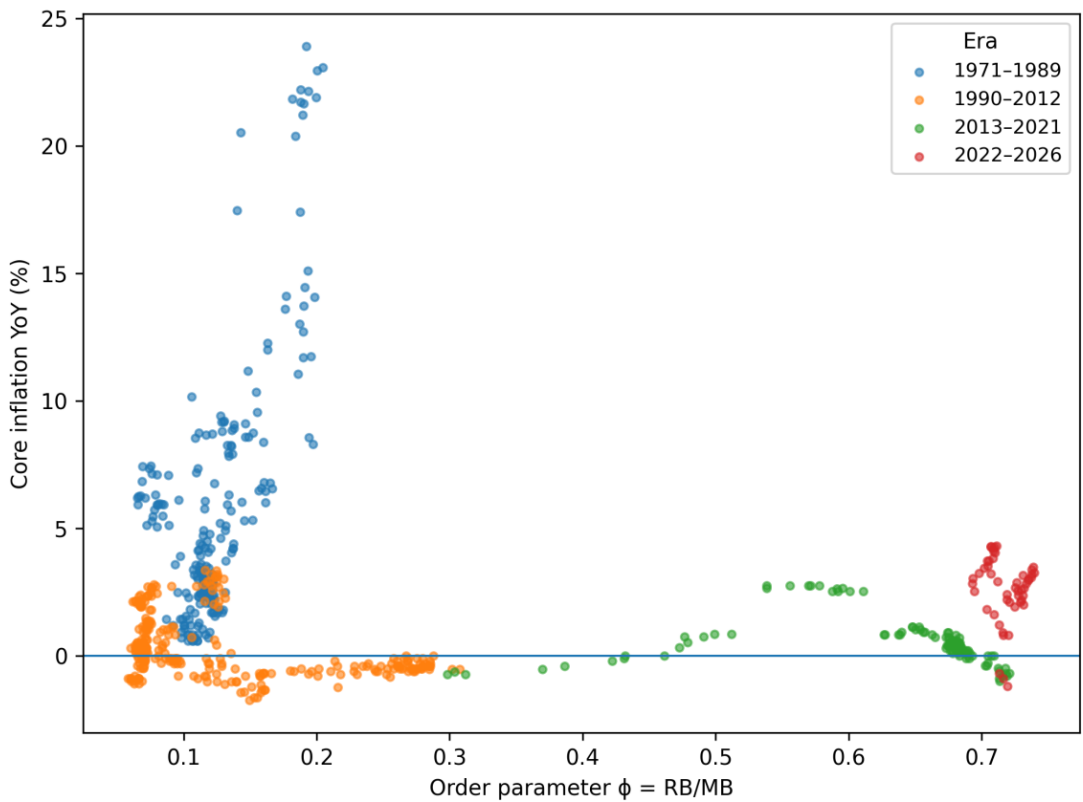

**Fig. 3. Phase diagram in $(\phi, \pi^{core})$ space.**

Monthly observations of core CPI inflation, $\pi^{core}$ (YoY, %), versus the order parameter $\phi$. Points are grouped by era (1971–1989, 1990–2012, 2013–2021, and 2022–2026). High-$\phi$ states (reserve-dominated phase) are compatible with both near-zero inflation (2013–2021) and the post-2022 inflation recovery.

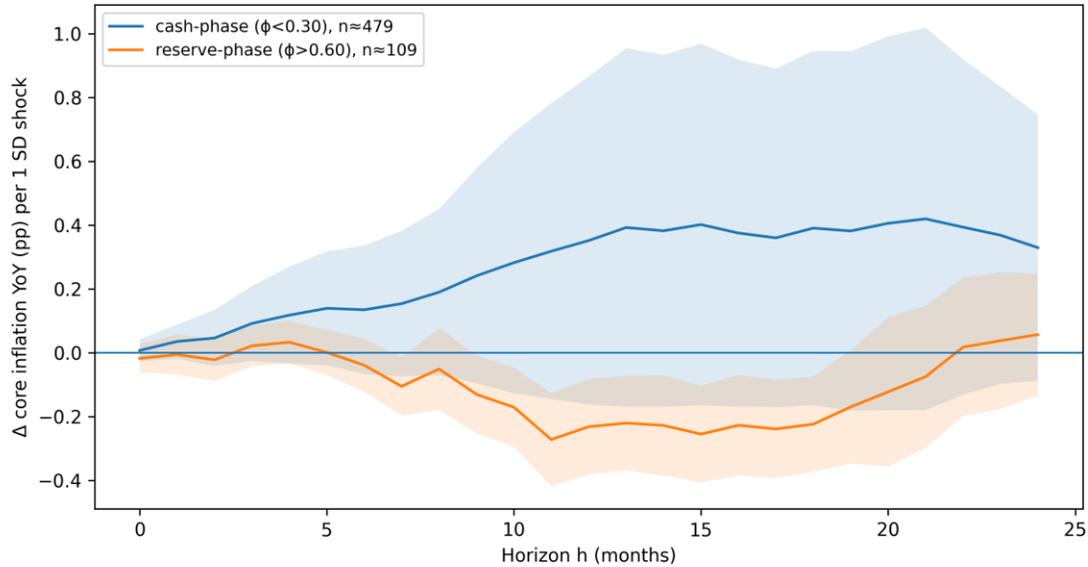

**Fig. 4. Phase-dependent inflation kernel.**
Phase-conditional local-projection impulse responses of core CPI inflation to a one-standard-deviation unexpected base-growth shock, $u_t^{MB}$. The shock is constructed as the within-phase AR(12) residual of year-on-year base-money growth and standardized within each phase. Cash phase: $\phi < 0.30$; reserve phase: $\phi > 0.60$; the intermediate critical region is excluded. Shaded bands indicate 95% HAC(12) confidence intervals. Local projections follow Jordà.

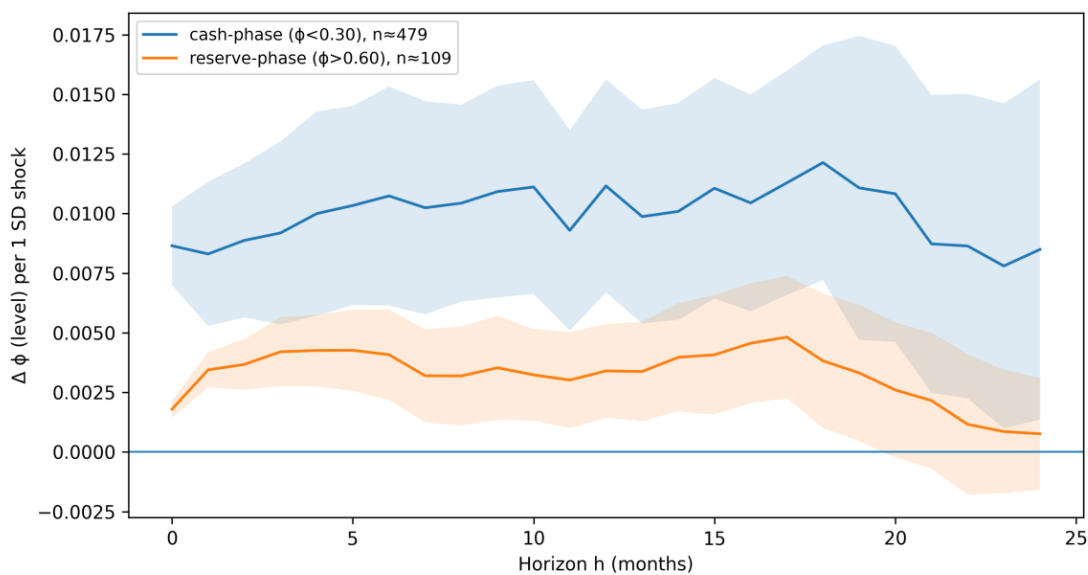

**Fig. 5. Reservoir trapping closes the mechanism loop.**
Phase-conditional local-projection impulse responses of the order parameter $\phi$ to the same one-standard-deviation unexpected base-growth shock, $u_t^{MB}$, used in Fig. 4. Responses are significantly positive in both phases, indicating that base-money expansions preferentially populate the reserve reservoir. Shaded bands indicate 95% HAC(12) confidence intervals.

# Supplementary Materials I

**for:** *A phase transition in monetary function explains expansion without inflation*

# Contents

- S1. Data sources and preprocessing
- S2. Breakpoints and robustness
- S3. Impulse-response specifications (local projections)
- S4. Full IRF tables (CSV)
- S5. Additional robustness checks
- Supplementary figure captions

---

# S1. Data sources and preprocessing

## S1.1 Data sources

We use two public monthly datasets.

**(i) Monetary base and components (Bank of Japan).**
We use the Bank of Japan long-term time series file *Monetary Base (Average amounts outstanding), long-term time series* (`mblong.xlsx`). The analysis uses the sheet corresponding to average amounts outstanding, which provides monthly series for the monetary base ($MB_t$), banknotes in circulation ($BN_t$), coins in circulation ($CO_t$), reserve balances ($RB_t$), and the seasonally adjusted monetary base ($MB_t^{SA}$). All monetary quantities are measured in 100 million yen. The date index is aligned at month-end.

**(ii) CPI (Statistics Bureau of Japan; e-Stat).**
We use monthly CPI time series (2020 base) from the Statistics Bureau of Japan (e-Stat). We extract headline CPI ("All items"), denoted $CPI_t$, and core CPI ("All items less fresh food and energy"), denoted $CPI_t^{core}$. The CPI series are index numbers with 2020 = 100 and are aligned at month-end. Category definitions follow the official statistical documentation.

## S1.2 Variable construction

All derived variables are constructed after merging the monetary and CPI series on the month-end date index.

**Order parameter (phase variable):**

$$\phi_t = \frac{RB_t}{MB_t}.$$

**Inflation measures (year-over-year):**

$$\pi_t = 100\left(\frac{CPI_t}{CPI_{t-12}} - 1\right), \qquad \pi_t^{core} = 100\left(\frac{CPI_t^{core}}{CPI_{t-12}^{core}} - 1\right).$$

**Monetary base growth (year-over-year, seasonally adjusted):**

$$g_t^{MB} = 100\left(\frac{MB_t^{SA}}{MB_{t-12}^{SA}} - 1\right).$$

The main analysis begins in 1971-01 because year-over-year variables require twelve prior months of data.

### S1.3 Indexing for level plots

For level-comparison plots (Fig. 1A), each level series $x_t$ is transformed into an indexed series:

$$x_t^{idx} = 100 \cdot \frac{x_t}{x_{t_0}},$$

where $t_0$ is the first month of the merged sample. These indexed series are plotted on a logarithmic scale.

### S1.4 Phase definitions and sample ranges

To estimate phase-conditioned response functions, we classify months into two regimes using the order parameter $\phi_t$:

- **cash phase:** $\phi_t < 0.30$
- **reserve phase:** $\phi_t > 0.60$
- **intermediate region:** $0.30 \leq \phi_t \leq 0.60$

The intermediate region is excluded from phase-conditional impulse-response estimation and is interpreted as a transition or critical region. The separation between the two thresholds is deliberate and reduces phase mixing in the local-projection estimates. Threshold sensitivity is examined in S5.

---

## S2. Breakpoints and robustness

### S2.1 Motivation

The main text identifies a compositional phase transition around 2013. Here we verify that this interpretation does not depend on a single window choice or a single prespecified break date.

We study breakpoints for three series:

1. $\log MB_t^{SA}$
2. $\phi_t$
3. $\pi_t^{core}$

These variables represent, respectively, the scale of the monetary base, the composition of the monetary base, and the core inflation outcome that the main text seeks to explain.

## S2.2 Automatic single-break detection

Within a sample window $[t_a, t_b]$, we estimate a single breakpoint $\tau$ by fitting a two-segment linear model and selecting the breakpoint that minimizes the residual sum of squares, subject to a minimum segment length of 24 months on each side:

$$y_t = \alpha_1 + \beta_1 t, \qquad t \leq \tau,$$
$$y_t = \alpha_2 + \beta_2 t, \qquad t > \tau.$$

We implement this procedure by grid search over all admissible candidate months within each window after excluding the minimum-length margins.

## S2.3 Window sets

We examine three breakpoint clusters corresponding to major regime periods:

- **1990 cluster:** five windows, spanning approximately 1983-1997 with nested tighter variants
- **2013 cluster:** five windows, spanning approximately 2008-2019 with nested tighter variants
- **2022 cluster:** four to five windows, spanning approximately 2018-2024/2025 with nested variants

The full list of windows is provided in the accompanying reproducibility code used to generate Fig. S1.

## S2.4 Results (Fig. S1)

Figure S1 reports the distribution of detected breakpoints across alternative windows for each series and each cluster.

For the **2013 cluster**, breakpoints in $\log MB_t^{SA}$ concentrate tightly in 2013, while breakpoints in $\phi_t$ cluster in the same transition zone. Breakpoints in core inflation tend to occur later, often around 2014. This pattern is consistent with a compositional transition followed by delayed consumer-price transmission.

For the **1990 cluster**, core inflation exhibits a stable breakpoint near the late-1980s to early-1990 transition, while $\phi_t$ shifts more clearly in the early 1990s, consistent with delayed compositional adjustment.

For the **2022 cluster**, breakpoints in core inflation are comparatively stable around late 2022, whereas breakpoints in $\log MB_t^{SA}$ and $\phi_t$ are more sensitive to window choice. This is consistent with a mixed period combining inflation recovery and policy normalization.

Taken together, these results support the main-text interpretation that the post-2013 episode reflects not only quantity expansion but also compositional reorganization within the monetary base.

---

## S3. Impulse-response specifications (local projections)

### S3.1 Overview

We estimate phase-conditioned impulse responses (IRFs) using local projections (LP). The main response variables are:

- $y_t = \pi_t^{core}$ for Fig. 4 and the associated supplementary robustness figures
- $y_t = \phi_t$ for Fig. 5 and the associated supplementary robustness figures

The shock variable is the unexpected monetary-base-growth shock, denoted $u_t^{MB}$.

### S3.2 Shock construction: within-phase AR residualization

A natural concern is that policy actions may respond endogenously to inflation conditions. To reduce predictable or reactive components, we construct shocks separately within each phase.

Let $x_t = g_t^{MB}$, the year-over-year growth rate of the seasonally adjusted monetary base. Within each phase-specific subsample, we estimate an AR(12) model:

$$x_t = a_0 + \sum_{i=1}^{12} a_i \, x_{t-i} + u_t.$$

The residual $u_t$ is defined as the unexpected component $u_t^{MB}$. We standardize $u_t^{MB}$ to unit variance within each phase so that the reported impulse responses represent responses to a one-standard-deviation shock.

This procedure is reduced-form rather than fully structural. Its purpose is to remove serially predictable components of monetary-base growth and to mitigate simple reaction-function contamination.

### S3.3 Local projection equation

For each horizon $h = 0, 1, \ldots, H$, with baseline horizon $H = 24$ months, we estimate

$$y_{t+h} = \alpha_h + \beta_h u_t^{MB} + \sum_{i=1}^{L} \gamma_{h,i}\, y_{t-i} + \sum_{i=1}^{L} \delta_{h,i}\, u_{t-i}^{MB} + \varepsilon_{t+h},$$

where the baseline lag order is $L = 12$. The coefficient $\beta_h$ is the impulse response at horizon $h$.

### S3.4 Standard errors and confidence intervals

We compute heteroskedasticity- and autocorrelation-consistent standard errors using a Newey-West HAC estimator with maximum lag 12. We report 95% confidence intervals as

$$\beta_h \pm 1.96 \cdot SE_h.$$

The shaded bands in the main-text and supplementary IRF figures correspond to these intervals.

### S3.5 Phase conditioning and sample construction

Local projections are estimated separately for the **cash phase** ($\phi_t < 0.30$) and the **reserve phase** ($\phi_t > 0.60$). The intermediate region is excluded from the baseline phase-conditioned regressions in order to avoid mixing distinct regimes in the estimated response kernels.

### S3.6 Notes on interpretation

The LP coefficients are reduced-form dynamic responses conditional on the specified controls and shock definition. We interpret them as phase-conditioned response kernels. The empirical claims supported by these estimates are limited and testable:

1. the response kernel differs across phases;
2. unexpected base-growth shocks are absorbed into reserves, raising $\phi_t$;
3. the qualitative contrast between phases is robust to reasonable variations in thresholds and specification (S5).

We do not claim a fully structural decomposition of the mechanisms underlying the sign and shape differences. Rather, we document a robust phase dependence consistent with the phase-transition interpretation developed in the main text.

---

## S4. Full IRF tables (CSV)

### S4.1 Provided files

We provide full IRF tables for the two main-text impulse-response figures:

- `IRF_J6_core_inflation.csv` for Fig. 4
- `IRF_J7_phi.csv` for Fig. 5

For each horizon and each phase, these files report:

- `beta`: estimated IRF coefficient
- `se`: HAC(12) standard error
- `ci_low`, `ci_high`: lower and upper bounds of the 95% confidence interval
- `n`: effective regression sample size

Metadata fields also identify the phase label, shock definition, response variable, horizon length, and lag length.

### S4.2 Reproducibility from CSV

To reproduce Fig. 4, filter `IRF_J6_core_inflation.csv` by phase and plot beta against horizon $h$ with a shaded band spanning `ci_low` to `ci_high`.

To reproduce Fig. 5, apply the same procedure to `IRF_J7_phi.csv`.

These tables make the quantitative impulse-response results directly inspectable and independently reproducible.

---

## S5. Additional robustness checks

### S5.1 Threshold sensitivity (Figs. S2-S3)

To assess whether the phase-dependent results are sensitive to the baseline phase classification, we re-estimate the phase-conditioned local projections under alternative threshold pairs. Specifically, we consider tighter and looser definitions around the baseline split, including cash-phase thresholds of $\phi_t < 0.25$ and $\phi_t < 0.35$, together with reserve-phase thresholds of $\phi_t > 0.65$ and $\phi_t > 0.55$.

The resulting IRFs are reported in **Fig. S2** for the core inflation response and in **Fig. S3** for the order-parameter response. Across these alternative threshold choices, the main qualitative pattern remains unchanged: the inflation-response kernel differs systematically between the two phases, and positive base-growth shocks remain associated with an increase in $\phi_t$, consistent with reservoir trapping.

### S5.2 Horizon and lag sensitivity (Fig. S4)

We next assess sensitivity to the local-projection specification itself by varying the horizon length $H$ and lag length $L$. In particular, we examine horizons of 12, 24, and 36 months together with lag orders of 6, 12, and 18 months.

The corresponding results are shown in **Fig. S4**. The medium-horizon contrast between the cash phase and reserve phase remains qualitatively stable across reasonable parameter choices. While the exact magnitude and smoothness of the estimated responses vary somewhat with specification, the phase dependence of the response kernel is preserved.

### S5.3 Alternative shock definitions (Fig. S5)

To verify that the results do not depend on a single residualization procedure, we consider alternative constructions of the monetary-base-growth shock. In addition to the baseline within-phase AR(12) residual, we examine specifications based on alternative autoregressive orders and a detrended-growth residualization.

Specifically, we consider:

1. AR residuals with alternative lag orders, including AR(6) and AR(18), in place of the baseline AR(12);
2. a detrended-growth residual in which $g_t^{MB}$ is regressed on a time trend and lagged terms within phase, and the residual is used as the shock.

The resulting IRFs are reported in **Fig. S5**. The reserve-phase response remains distinct from the cash-phase response, and the positive response of $\phi_t$ to unexpected base-growth shocks is preserved. These results indicate that the main findings are not an artifact of one particular shock definition.

### S5.4 Critical-region diagnostic

As an additional diagnostic, we examine the intermediate region $0.30 \leq \phi_t \leq 0.60$ separately. Estimates in this region are less stable than those in the two outer regimes, with wider confidence intervals and greater sensitivity to specification. This behavior is consistent with interpreting the intermediate interval as a transition region rather than a stable phase.

### S5.5 Summary of robustness outcomes

Taken together, the supplementary robustness checks support three qualitative conclusions:

1. the post-2013 episode is robustly associated with a compositional shift in $\phi_t$;
2. positive monetary-base shocks are absorbed into reserves, raising $\phi_t$ rather than transmitting proportionally into prices;
3. the inflation-response kernel differs systematically between the cash phase and reserve phase across a range of reasonable thresholds and specifications.

# Supplementary figure captions

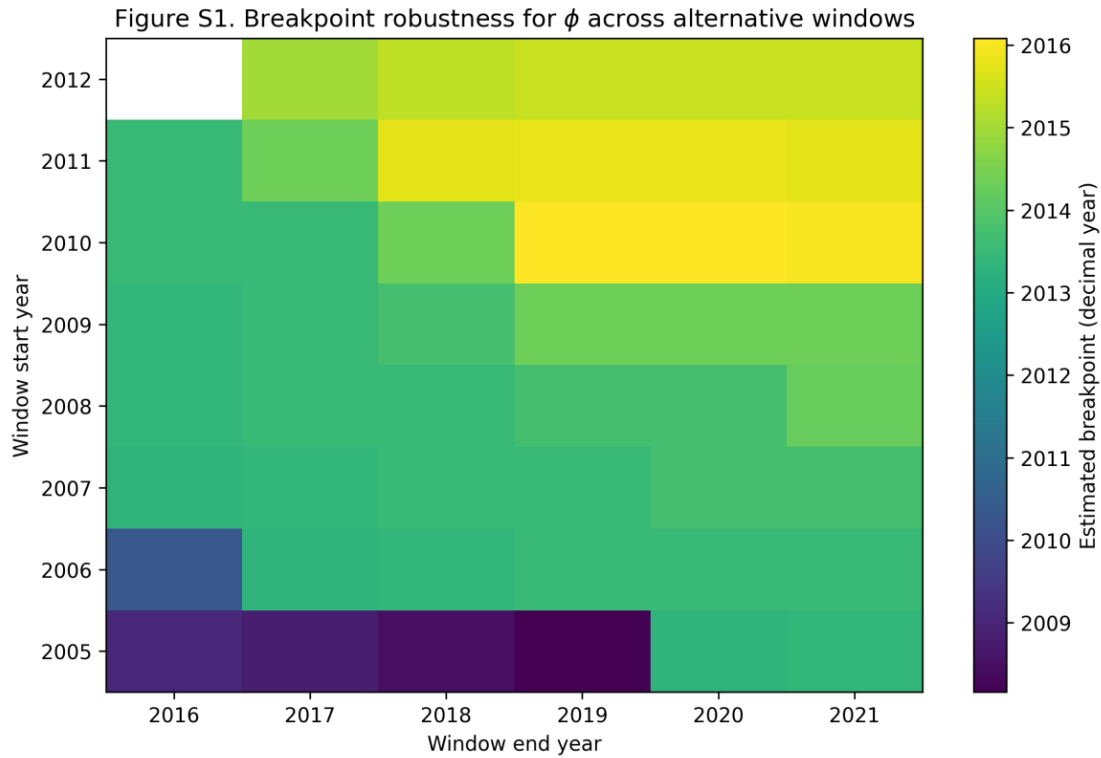

**Fig. S1. Breakpoint robustness across alternative windows (automatic detection).**
Automatic single-break estimates are obtained by minimizing the residual sum of squares of a two-segment linear model, subject to a minimum segment length of 24 months, across multiple window choices around the 1990, 2013, and 2022 clusters. Panels report detected breakpoints for $\log(MB_t^{SA})$, $\phi_t$, and core inflation measured year over year.

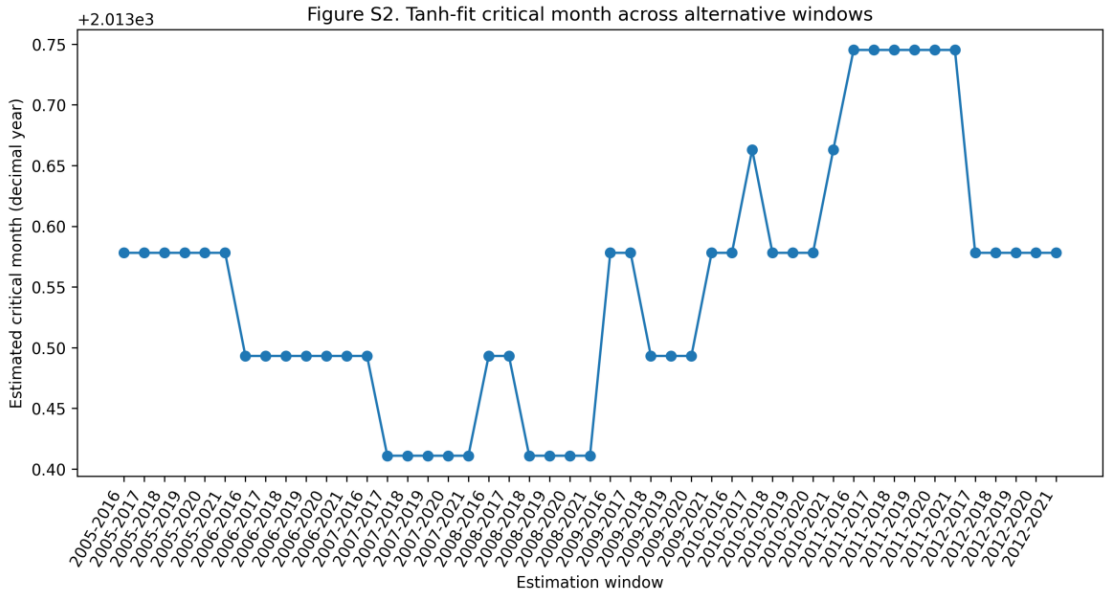

**Fig. S2. Threshold sensitivity of the phase-conditioned core inflation response.** Local-projection impulse responses of core inflation to an unexpected monetary-base-growth shock under alternative phase thresholds. The qualitative contrast between cash-phase and reserve-phase responses remains visible across tighter and looser classifications.

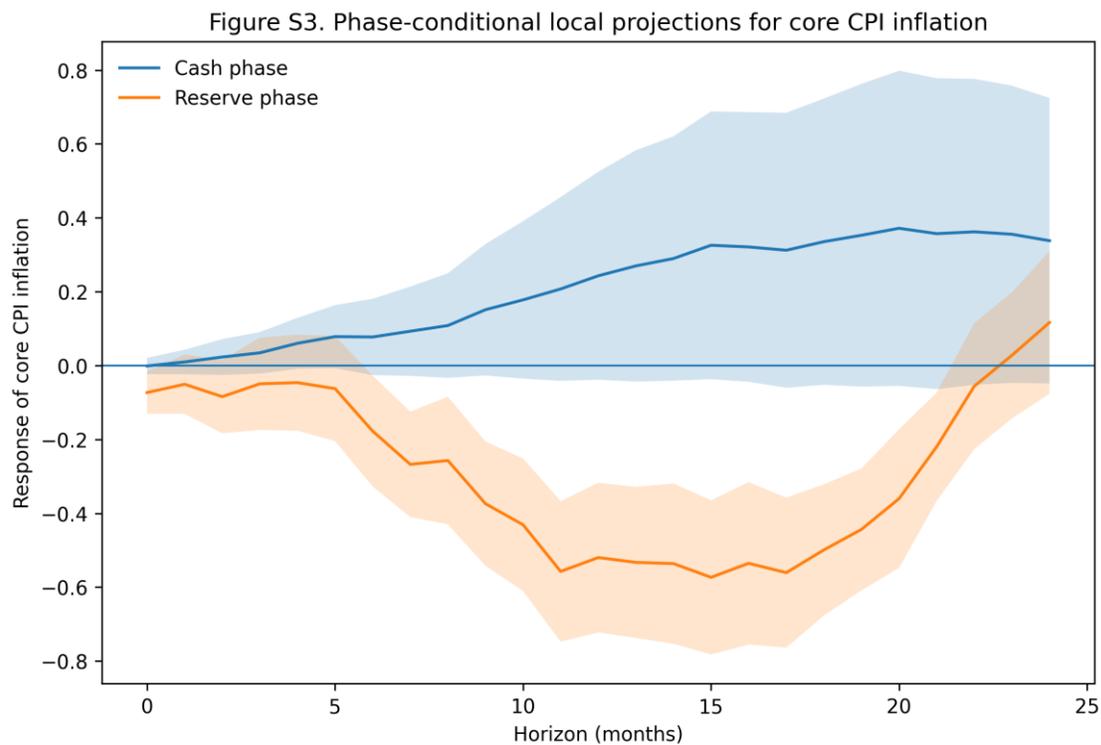

**Fig. S3. Threshold sensitivity of the order-parameter response.** Local-projection impulse responses of the order parameter $\phi_t$ to an unexpected monetary-base-growth shock under alternative phase thresholds. Across

specifications, positive shocks continue to raise $\phi_t$, consistent with reserve absorption.

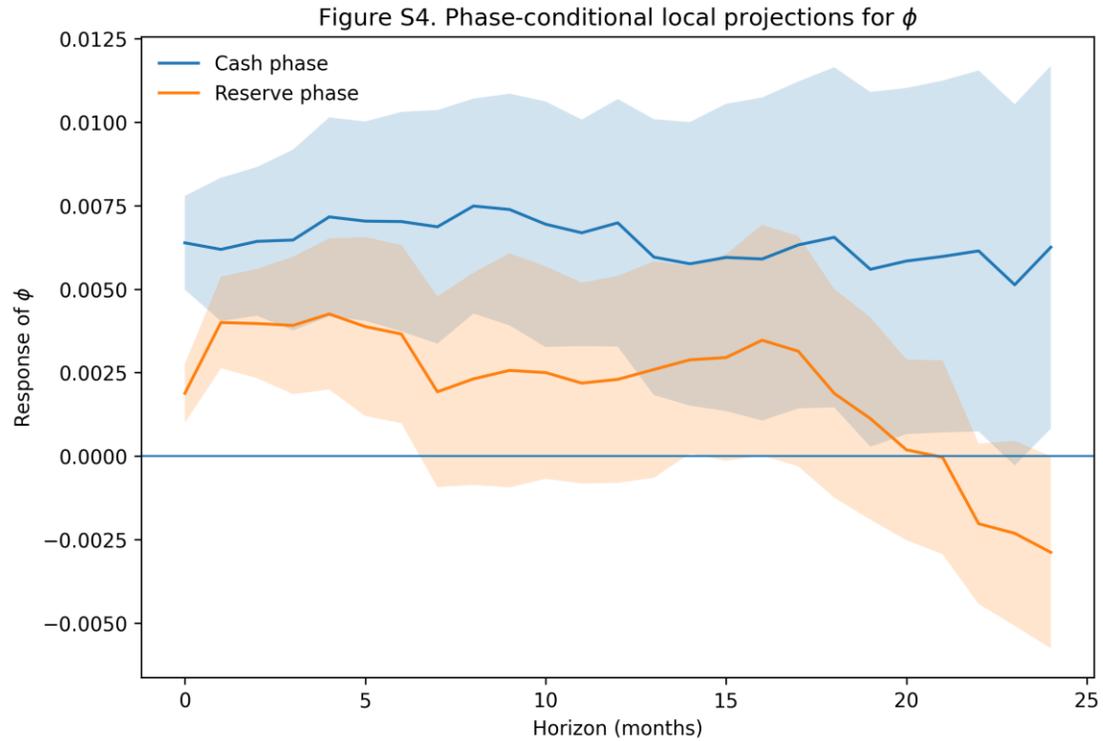

**Fig. S4. Horizon and lag sensitivity of the phase-conditioned local projections.** Impulse responses re-estimated under alternative horizon lengths and lag orders. Although magnitudes vary somewhat across specifications, the medium-horizon phase contrast remains qualitatively stable.

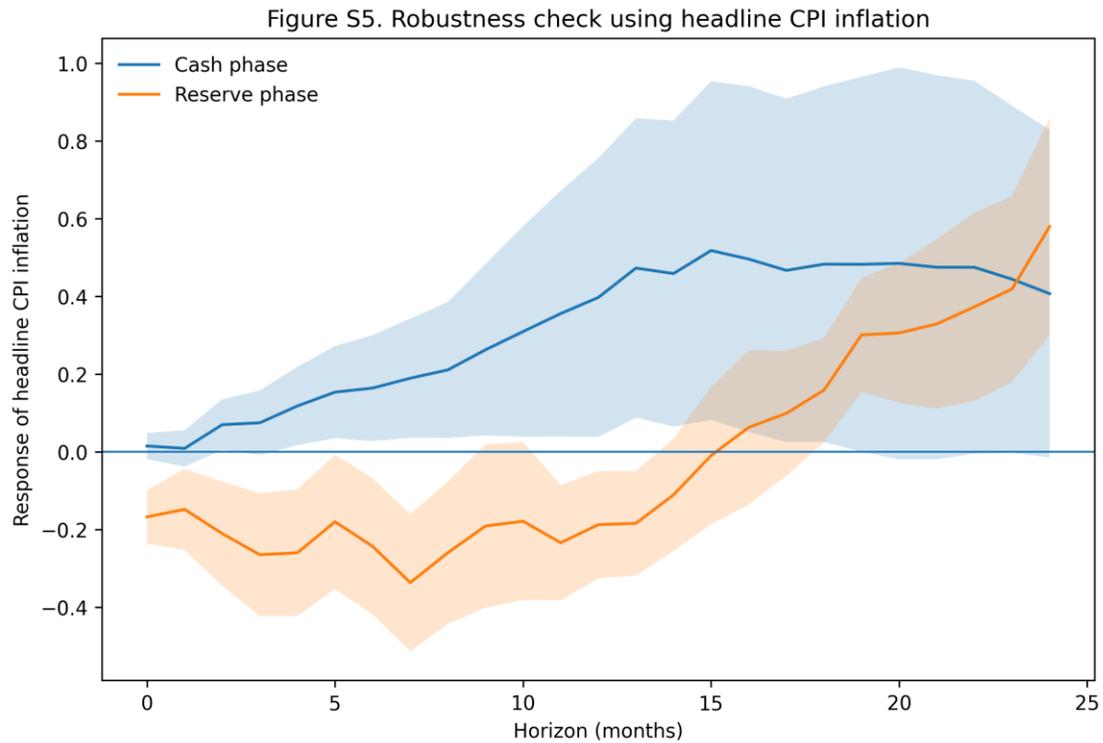

**Fig. S5. Alternative shock-definition robustness.**
Impulse responses obtained using alternative constructions of the monetary-base-growth shock, including different autoregressive residualizations and detrended-growth residuals. The main phase-dependent patterns remain qualitatively unchanged.

# Supplemental II: The Physics Modelling and Analysis

## S2.1 Overview

This supplement develops a standalone physical modelling framework for the Japan case. The purpose is not to redescribe the empirical findings in metaphorical language, but to construct a minimal calculable model that can define an order parameter, generate a phase transition, identify a critical point, and explain why the CPI transmission kernel can weaken or reverse sign once the system becomes reserve-dominated. The framework is built in three layers. First, a dissipative two-compartment model is introduced, consisting of a reservation reservoir and a circulation-coupled liquidity pool. Second, the model is coarse-grained into a Landau free-energy description for the order parameter. Third, a constitutive CPI coupling law is introduced with a critical order parameter $\phi_c$, allowing the sign of the inflation response to change across phases.

## S2.2 First-principles starting point

The starting point is the SCR allocation identity

$$G = S + C + R,$$

where $S$ denotes reproduction, $C$ contemporaneous consumption, and $R$ reservation or buffer storage. The first-principles claim is that $R$ is a dissipative compartment. In the original formulation this can be interpreted as physical deterioration; in a modern balance-sheet setting it appears as effective carrying costs, liquidity regulation, precautionary absorption, and intermediation friction. In such a system, an external symbolic input is expected to compensate or fill the dissipative buffer before it amplifies the consumption channel. This is the physical core of SRF priority.

## S2.3 Minimal two-compartment dynamics

We define two state variables: $R(t)$, the reservation reservoir, and $X(t)$, the circulation-coupled liquidity. For an impulse injection at $t = 0$, the model is

$$\dot{R} = -\delta R + \eta X, \qquad \dot{X} = -\gamma X,$$

with initial conditions

$$R(0^+) = A, \qquad X(0^+) = B.$$

Here $A$ is the directly absorbed reservoir share of the impulse, $B$ is the directly circulation-coupled share, $\delta$ is the effective reservoir relaxation rate, $\gamma$ is the circulation decay rate, and $\eta$ is the reabsorption rate from $X$ back into $R$. The closed-form impulse responses are

$$X(h) = Be^{-\gamma h},$$

$$R(h) = Ae^{-\delta h} + \eta B \frac{e^{-\gamma h} - e^{-\delta h}}{\delta - \gamma}.$$

This is already a genuinely calculable dynamical system rather than a descriptive overlay.

## S2.4 Order parameter and linearization

Define the order parameter

$$\phi = \frac{R}{R + X}.$$

Linearizing around a phase-specific baseline $\bar{\phi}$ yields

$$\Delta\phi(h) \propto (1 - \bar{\phi})\Delta R(h) - \bar{\phi}\Delta X(h).$$

This relation is geometric. Reservoir increments raise $\phi$, while circulation increments lower it, with weights determined by the background phase. It provides the minimal bridge between the two-compartment state dynamics and the empirically measured reserve-share fraction.

## S2.5 Calibration to Japan

Using phase-specific local-projection IRFs from Japan monthly data (1971–2026), the model is fitted separately in the cash phase ($\phi < 0.30$) and the reserve phase ($\phi > 0.60$). The fitted parameter table is provided in `two_compartment_parameters.csv`, and the critical-point summary is in `critical_point_summary.csv`. The fitted model reproduces the qualitative structure of the empirical IRFs: reservoir trapping remains positive in both phases, while the CPI response is weakly positive in the cash phase but negative in the reserve phase.

Figure S2-1 compares the empirical and model-implied impulse responses in the cash phase. The key point is that the same minimal dynamical structure can simultaneously account for a positive $\Delta\phi$ response and a weakly positive CPI response in the low-$\phi$ regime.

Figure S2-2 shows the corresponding comparison in the reserve phase. Here the model captures the distinctive pattern that motivates the entire phase-based framework: reservoir trapping remains positive, but the CPI response becomes attenuated and turns negative. This is the dynamical signature of a reserve-dominated regime.

## S2.6 Effective CPI coupling and the critical point

To connect the model to prices, we use the constitutive law

$$\Delta \pi(h) = s_\pi \, \chi(\bar{\phi}) \, \Delta X(h),$$

with

$$\chi(\bar{\phi}) = 1 - \frac{\bar{\phi}}{\phi_c}.$$

The parameter $\phi_c$ is the critical order parameter at which the effective CPI coupling changes sign. In the current Japan calibration, the fitted critical point is approximately

$$\phi_c \approx 0.231.$$

For comparison, the empirical phase means are approximately

$$\bar{\phi}_{\text{cash}} \approx 0.127, \qquad \bar{\phi}_{\text{reserve}} \approx 0.694.$$

Therefore

$$\bar{\phi}_{\text{cash}} < \phi_c < \bar{\phi}_{\text{reserve}},$$

which is exactly the ordering required for a sign change in the inflation kernel across phases.

Figure S2-3 visualizes the effective coupling function $\chi(\phi)$. The figure is important because it shows that the cash and reserve phases lie on opposite sides of the same critical point. The sign flip of CPI transmission is therefore not imposed ad hoc for each regime separately, but arises from a single common constitutive threshold.

## S2.7 Why a sigmoidal transition appears

Suppose the direct reservoir absorption share follows a logistic control law

$$a(\theta) = \frac{1}{1 + e^{-\lambda(\theta - \theta_c)}}.$$

Then the steady-state order parameter $\phi^*(\theta)$ becomes sigmoidal, because reservoir occupation rises nonlinearly as direct absorption into $R$ becomes dominant. This provides the mechanical underpinning for the tanh/logistic transition observed in the empirical order parameter.

Figure S2-4 shows the resulting steady-state phase trajectory. Its role is to demonstrate that the observed tanh-like empirical transition does not need to be introduced as a purely phenomenological fit; it can be generated from a minimal microscopic allocation law.

## S2.8 Landau coarse-graining

The two-compartment model uses two state variables:

- $R(t)$: reservation reservoir
- $X(t)$: circulation-coupled liquidity

with order parameter

$$\phi(t) = \frac{R(t)}{R(t) + X(t)}.$$

At the dynamical level, the model already gives calculable impulse responses. To move from a response model to a genuine phase-transition model, we coarse-grain the system and retain only the slow order parameter $\phi$. The fast balance-sheet details are absorbed into an effective free-energy landscape. Let the shifted order parameter be

$$m = \phi - \phi_c.$$

This is the natural Landau variable: the transition occurs around $m = 0$, not around $\phi = 0$. We then define an effective free energy

$$F(m; \theta) = \frac{1}{2}a(\theta)m^2 + \frac{1}{4}bm^4 - hm,$$

where $\theta$ is a control parameter, $a(\theta)$ is the control-dependent quadratic coefficient, $b > 0$ stabilizes the system, and $h$ is an external bias field. The over-damped Landau–Khalatnikov dynamics is

$$\tau \dot{m} = -\frac{\partial F}{\partial m} + \zeta(t),$$

or equivalently

$$\tau \dot{m} = -a(\theta)m - bm^3 + h + \zeta(t).$$

where $\tau$ is the relaxation timescale and $\zeta(t)$ is noise.

The steady-state condition is therefore

$$a(\theta)m + bm^3 = h.$$

This equation determines the equilibrium branch $m^*(\theta)$, hence $\phi^*(\theta) = m^*(\theta) + \phi_c$. This is the formal step that lifts the two-compartment model from a purely dynamical description to an effective phase-transition theory. In this sense, the earlier tanh-like transition is not merely descriptive; it is the natural coarse-grained signature of a control parameter moving the system across a critical point.

Figure S2-5 shows the effective potential on the two sides of the critical point. Once $a(\theta)$ changes sign, the shape of the potential changes, and the system moves into a different dynamical basin. This is the potential-level explanation of the observed phase reorganization.

## S2.9 Critical susceptibility

The Landau picture also explains why the critical region is the most sensitive. Near the phase boundary, the restoring curvature of the effective potential becomes small, so the susceptibility rises sharply. A simple normalized representation is

$$\mathcal{S}(\phi) \propto \frac{1}{|\phi - \phi_c| + \varepsilon},$$

which peaks near $\phi_c$. This captures the physical meaning of criticality: the system responds most strongly to perturbations when it is close to the phase boundary.

Figure S2-6 makes this intuition explicit. The critical region is not merely the midpoint between two empirical regimes; it is the point at which the system becomes maximally responsive because the effective restoring force is weakest. ## S2.10. Connection to CPI sign reversal In the two-compartment formulation, CPI response is modeled through an effective coupling factor $\chi(\phi)$ that changes sign at $\phi_c$.
In the Landau formulation, this same critical point appears as the natural expansion point of the order parameter.

Hence the logic is unified:

1. the system is driven by a control parameter $\theta$,
2. the effective potential $F(m;\theta)$ changes shape across threshold,
3. the equilibrium order parameter crosses $m = 0$,
4. crossing $m = 0$ means $\phi$ crosses $\phi_c$,
5. once $\phi > \phi_c$, the CPI coupling kernel can weaken or reverse sign.

The physical significance of the critical point is that it also governs the sign of CPI transmission. The effective CPI coupling is positive when $\bar{\phi} < \phi_c$ and negative when $\bar{\phi} > \phi_c$. The Japan calibration therefore supports a unified interpretation: the cash phase and reserve phase do not require two unrelated mechanisms, but lie on opposite sides of the same critical point. In this framework, sign reversal of the inflation response is a direct consequence of phase-conditioned transmission.

## S2.11 What this supplement establishes

Taken together, this supplement turns the Japan mechanism into a compact physical theory rather than a descriptive overlay. It provides a dissipative two-compartment dynamical model, closed-form impulse responses, a calculable critical point $\phi_c$, a constitutive law for CPI sign reversal, a Landau effective-potential interpretation, and a susceptibility picture explaining why the critical region is the most sensitive. In this form, the "money supply does not necessarily cause inflation" result is no longer only an empirical statement; it becomes the consequence of a phase-dependent dynamical system in which newly created nominal capacity can be absorbed first by a dissipative reservation reservoir rather than by the circulation-coupled price-forming sector.

## S2.12 Figures

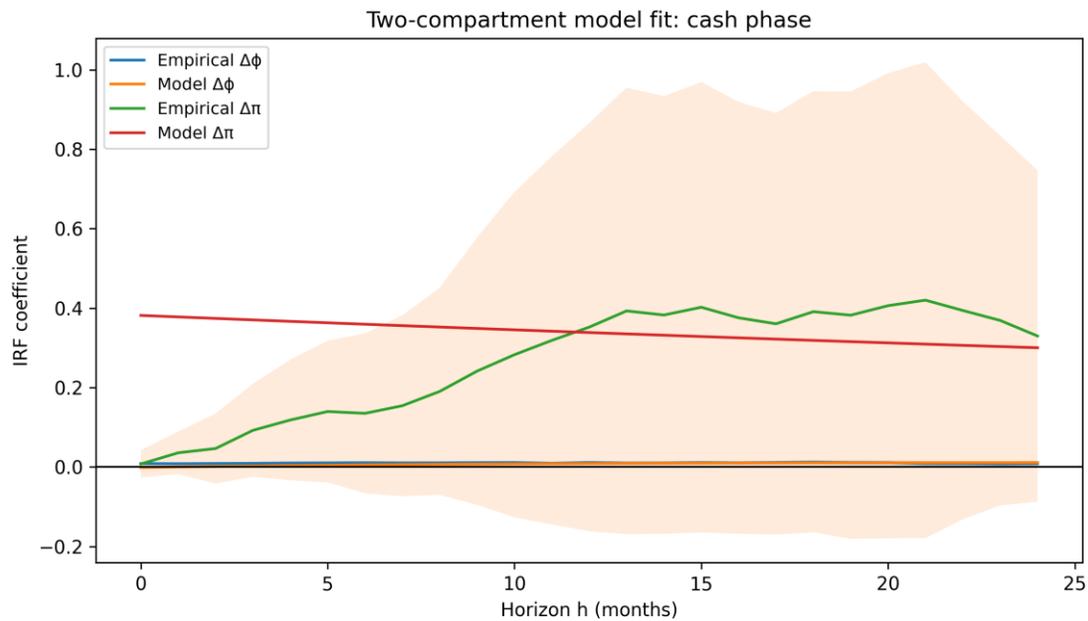

**Fig. S2-1. Empirical versus model IRFs in the cash phase.**
The fitted two-compartment model reproduces the qualitative structure of the cash-phase impulse responses: a positive order-parameter response and a weakly positive CPI response.

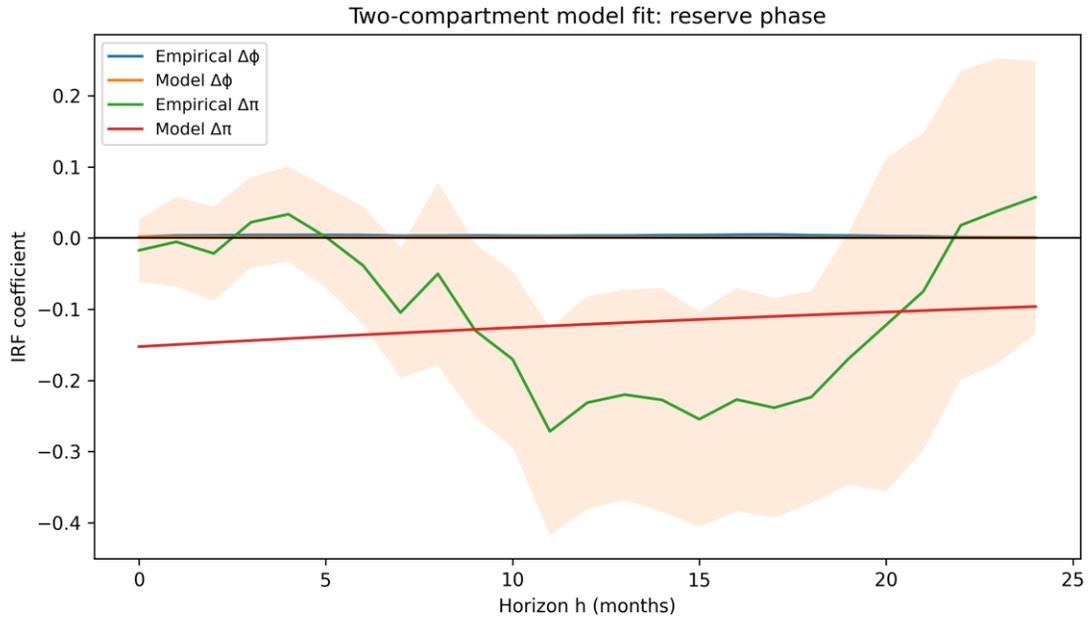

**Fig. S2-2. Empirical versus model IRFs in the reserve phase.**
The fitted two-compartment model reproduces the reserve-phase pattern in which reservoir trapping remains positive while the CPI response is attenuated and turns negative.

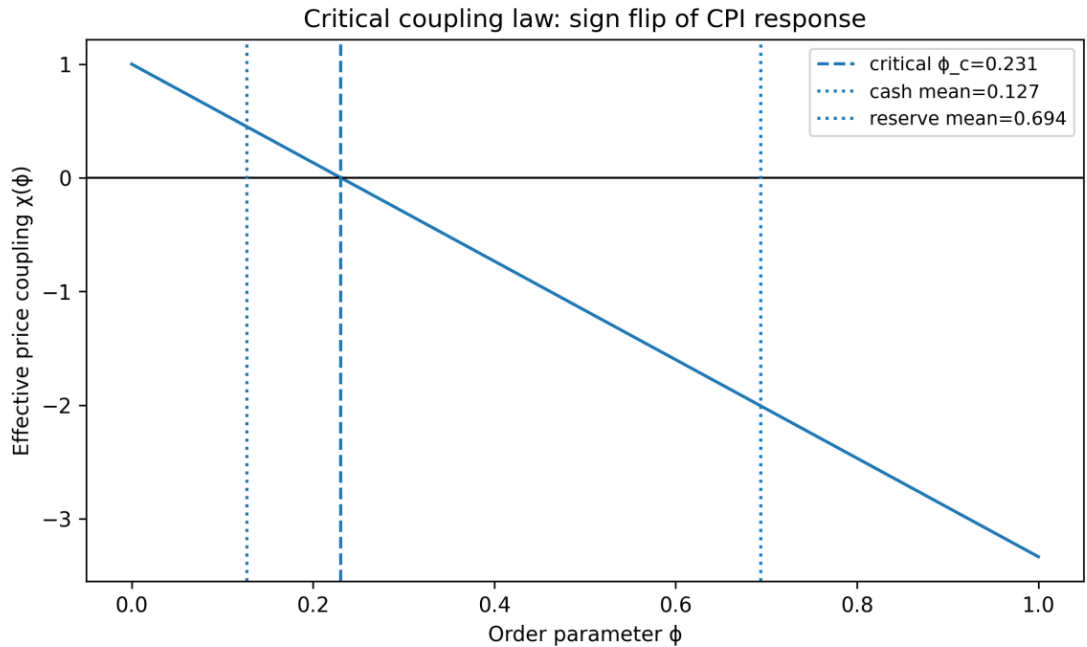

**Fig. S2-3. Critical coupling function and fitted critical point.**
The effective CPI coupling function $\chi(\phi) = 1 - \phi/\phi_c$ is shown together with the fitted critical point and the empirical mean order parameters of the cash and reserve phases.

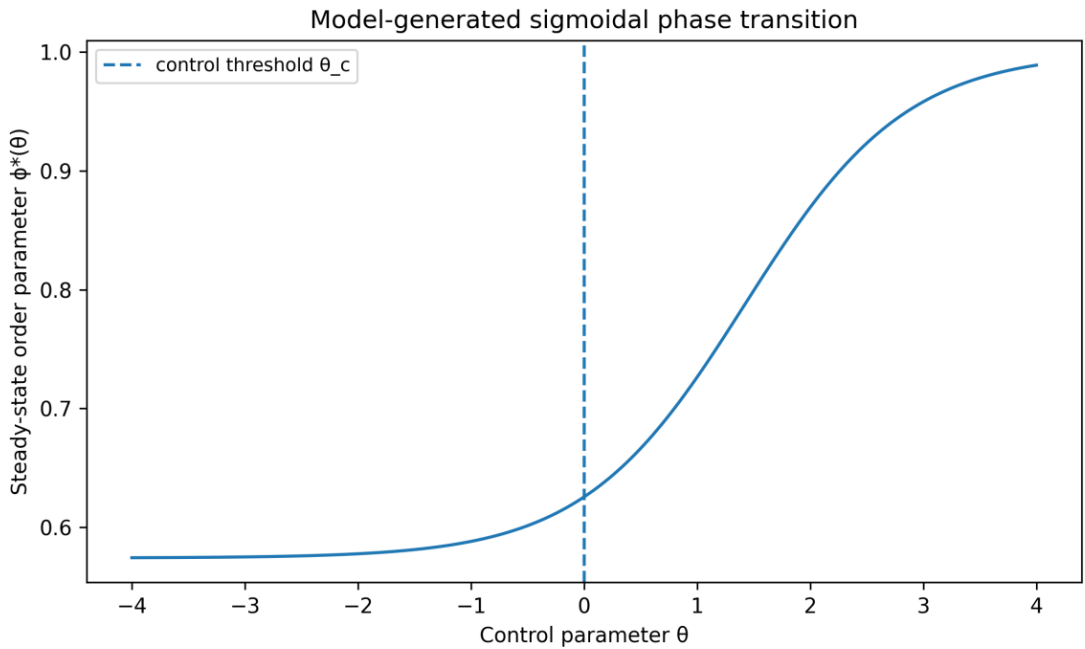

**Fig. S2-4. Model-generated sigmoidal steady-state phase transition.**
A logistic control law for reservoir absorption generates a sigmoidal steady-state order-parameter trajectory $\phi^*(\theta)$, providing a mechanistic basis for the empirical tanh-like transition.

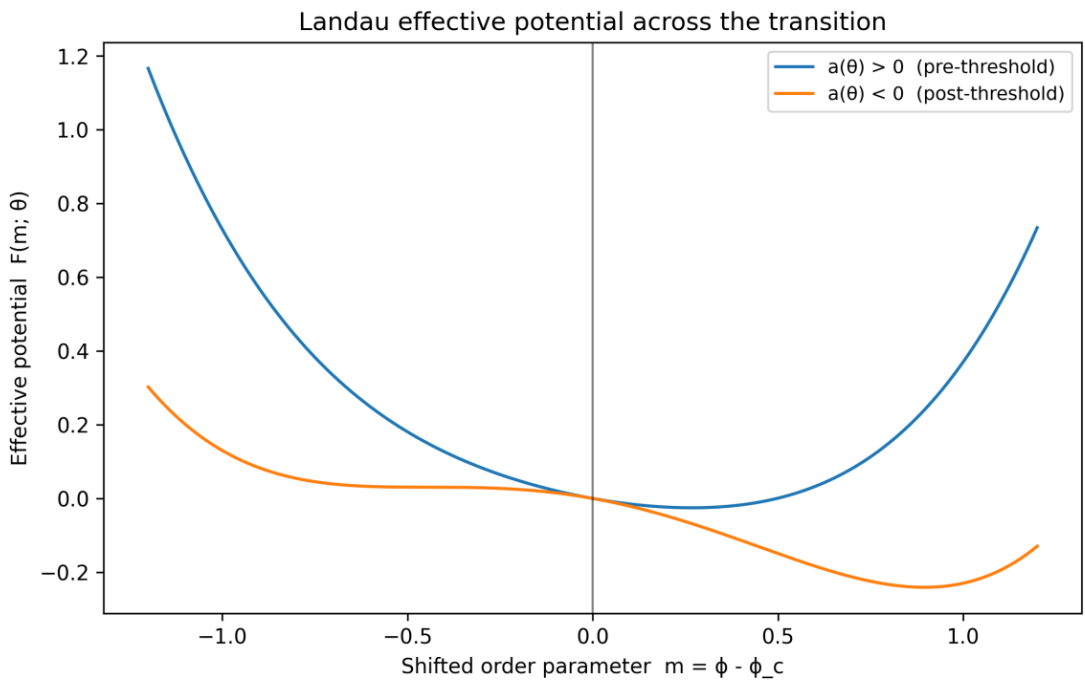

**Fig. S2-5. Landau effective potential across the transition.**
The effective free energy changes shape when the control parameter crosses threshold, illustrating the potential-level origin of the observed phase reorganization.

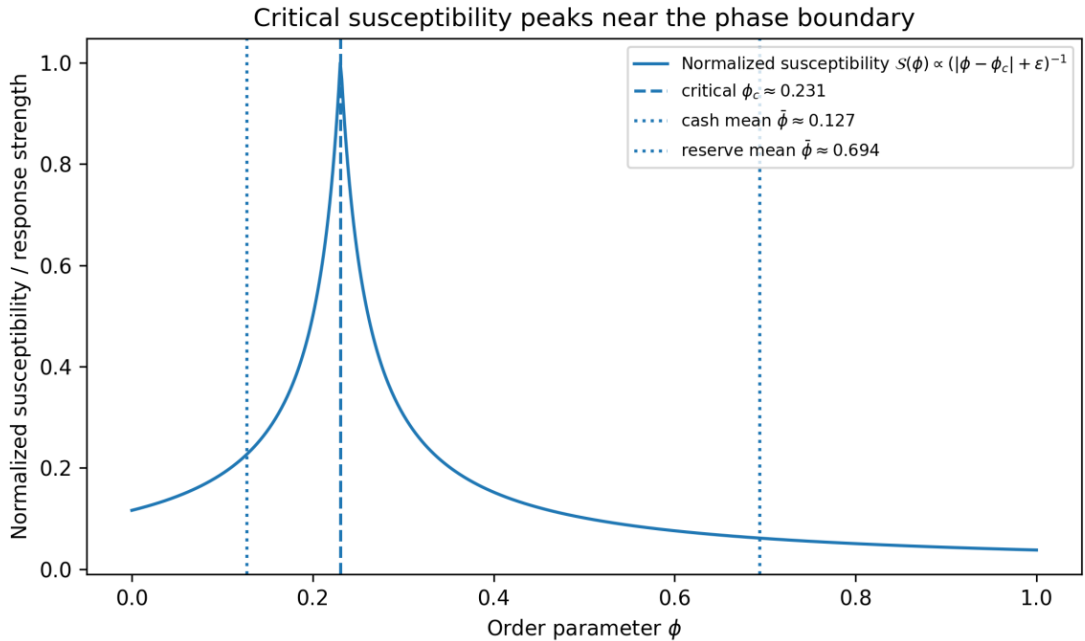

**Fig. S2-6. Critical susceptibility near the phase boundary.**
The response function peaks near $\phi_c$, showing that the critical region is the point of maximum sensitivity to perturbation.

## S2.12 Files associated with this supplement

The following files belong to this technical package:

- two_compartment_parameters.csv
- critical_point_summary.csv
- fit_cash_phase.csv
- fit_reserve_phase.csv

These files provide the intermediate theory notes, fitted parameter tables, phase-specific model fits, critical coupling visualization, steady-state phase transition figure, Landau potential, and critical susceptibility figure.